\documentclass[aps,prl,twocolumn,superscriptaddress]{revtex4-1}

\usepackage{graphicx}
\usepackage{bm}
\usepackage{rotating}
\usepackage{amssymb,amsmath}
\usepackage{multirow}
\usepackage{color}
 
\begin{document}

\title{Resonant X-ray scattering and the $j_{\mathrm{eff}}=1/2$ electronic
ground state in iridate perovskites}

\author{M. Moretti Sala}
\email{marco.moretti@esrf.fr}
\affiliation{European Synchrotron Radiation Facility, BP 220, F-38043 Grenoble
Cedex, France}

\author{S. Boseggia}
\affiliation{London Centre for Nanotechnology and Department of Physics and
Astronomy, University College London, London WC1E 6BT, United Kingdom}
\affiliation{Diamond Light Source Ltd, Diamond House, Harwell Science and
Innovation Campus, Didcot, Oxfordshire OX11 0DE, United Kingdom}

\author{D. F. McMorrow}
\affiliation{London Centre for Nanotechnology and Department of Physics and
Astronomy, University College London, London WC1E 6BT, United Kingdom}
\affiliation{Department of Physics, Technical University of Denmark, DK-2800
Kgs. Lyngby, Denmark}

\author{G. Monaco}
\affiliation{European Synchrotron Radiation Facility, BP 220, F-38043 Grenoble
Cedex, France}

\begin{abstract}
The resonant X-ray  scattering (magnetic elastic, RXMS, and inelastic, RIXS) 
of Ir$^{4+}$ at the L$_{2,3}$ edges relevant to 
spin-orbit Mott insulators  A$_{n+1}$Ir$_{n}$O$_{3n+1}$ (A=Sr, Ba, etc.) are
calculated using a single-ion model which treats the spin-orbit and tetragonal
crystal-field terms on an equal footing. Both RXMS and RIXS in the spin-flip
channel are found to display a non-trivial dependence on the direction of the
magnetic moment, $\boldsymbol\mu$. Crucially, we show that for  $\boldsymbol\mu$
in the \emph{ab}-plane,  RXMS at the L$_2$ edge is zero \emph{irrespective} of
the tetragonal crystal-field; spin-flip RIXS, relevant to measurements of
magnons, behaves reciprocally being zero at L$_2$ when $\boldsymbol\mu$ is 
perpendicular to the \emph{ab}-plane. Our results provide important insights
into the interpretation of X-ray data from the iridates, including that a
$j_{\mathrm{eff}}=1/2$ ground state cannot be assigned on the basis of
L$_2$/L$_3$ intensity ratio alone.
\end{abstract}

\maketitle

The existence of a Mott-like insulating ground state for specific members of the
Ruddleseden-Popper series of iridate perovskites A$_{n+1}$Ir$_{n}$O$_{3n+1}$
(A=Sr, $n=1,2$; A=Ba, $n=1$; A=Ca, $n=\infty$) has stimulated intense
interest\cite{Kim2008,Moon2008,Kim2009,Jackeli2009,Haskel2012,Ohgushi2013,
Boseggia2013}. Common wisdom had held that metallic ground states should be
displayed ubiquitously by 5d compounds due to the weakening of the onsite
Coulomb repulsion, $U$, and the broadening of the bandwidth, $W$, both resulting
from the extended nature of the 5d orbitals. It has been proposed, however, that
for Ir$^{4+}$ (5d$^5$) the strong spin-orbit coupling (SOC) present produces a
$j_{\mathrm{eff}}=1/2$ groundstate upon which even a moderate $U$ can act to
open a gap, hence leading to the formation of an insulating
state\cite{Kim2008,Moon2008}. While such spin-orbit Mott insulators are of
interest in their own right, further impetus for their study comes from
their structural similarities to certain cuprates, and not least the prediction
that they may form new families of superconductors\cite{Watanabe2010,Wang2011}.

Since the $j_{\mathrm{eff}}=1/2$ groundstate is actually an idealisation --
realised in perfect cubic symmetry only -- of pivotal importance is the need to
understand the robustness of this state to non-cubic structural distortions
found in real materials. This issue has been addressed through various
experimental probes, including optical absorption, angle-resolved photo
emission, X-ray absorption, etc.\cite{Kim2008,Moon2008} One technique that has
played a particularly prominent role in this endeavour is  resonant X-ray
magnetic scattering (RXMS), following the seminal work of Kim et
al.\cite{Kim2009} on Sr$_2$IrO$_4$ who argued that the near vanishing
of the RXMS intensity at the L$_2$ edge first observed in their experiments was
directly related to the existence of a $j_{\mathrm{eff}}=1/2$ ground state.
Although doubts have been raised concerning this
interpretation\cite{Chapon2011,Haskel2012}, others have followed the spirit of
Kim et al. and invoked the L$_2$/L$_3$ RXMS intensity ratio as a proxy for the
full understanding of the electronic
structure\cite{Boseggia2012,Boseggia2012a,JWKim2012,Calder2012,Boseggia2013,
Ohgushi2013}. This has lead to some unexpected conclusions, including the fact
that a $j_{\mathrm{eff}}=1/2$ groundstate is apparently realised in
Ba$_2$IrO$_4$\cite{Boseggia2013} even though the IrO$_6$ octahedra have a
tetragonal distortion almost twice as large as that in
Sr$_2$IrO$_4$\cite{Okabe2011}. Moreover, in bilayer Sr$_3$Ir$_2$O$_7$ the
magnetic moments undergo an unusual reorientation transition to point
perpendicular to the basal plane order displayed by the $n=1$ ``214''
counterparts while at the same time they display a L$_2$/L$_3$ RXMS intensity
ratio no larger than that of  the $n=1$
compounds\cite{Boseggia2012a,JWKim2012,Fujiyama2012}.

There is thus a clear need to elucidate fully the relationship between
the L$_2$/L$_3$ RXMS intensity ratio, the direction of the magnetic moment, and
the presence or otherwise of a $j_{\mathrm{eff}}=1/2$  ground state. To this end
we utilise a single-ion model which allows us to treat the SOC
$\zeta$ and a tetragonal crystal field $\Delta$ on an equal
footing\cite{Ament2011a,Liu2012,Ohgushi2013,Hozoi2012}. This model
has been chosen for the direct physical insight it provides. We use it to
explore both RXMS, and the RIXS in the spin-flip channel. This latter channel
has recently been exploited in various iridates to yield full magnon dispersion
curves across the entire Brillouin zone\cite{Kim2012,JKim2012}: information that
was previously the exclusive province of neutron spectroscopy. We focus in
particular on the explicit dependence of the X-ray scattering on the direction
of the local Ir$^{4+}$ magnetic moment, $\boldsymbol\mu$.  Realistic values of
$\zeta$ and $\Delta$ are chosen to aid comparison with experimental
data\cite{Liu2012,Gretarsson2013}. Our main finding is that both the RXMS and
RIXS in the spin-flip channel display a non-trivial dependence on the direction
of $\boldsymbol\mu$. We show that the L$_2$ edge RXMS intensity is identically
zero for magnetic moments lying in the
\emph{ab}-plane, \emph{irrespective} of the tetragonal crystal field splitting
of the t$_{2g}$ states, in agreement with the symmetry arguments of Ref.\
\cite{Chapon2011}. This has the important consequence that RXMS cannot be used
in isolation to deduce the existence or otherwise of the $j_{\mathrm{eff}}=1/2$
groundstate, calling for a reinterpretation of RXMS data on the  A$_2$IrO$_4$
(A=Sr and Ba) compounds  for which the magnetic moments are known to lie in the
\emph{ab}-plane\cite{Kim2009,Boseggia2013}. Our results are discussed with
reference to existing experimental data, and consideration given to their
implications for future work.

The calculation method adopted here for  Ir$^{4+}$ follows 
along similar lines to that in Refs. \onlinecite{GhiringhelliPhD,
MorettiSala2011} for L$_{2,3}$ edge RIXS in Cu$^{2+}$ cuprates (one-hole e$_g$
systems). For Ir$^{4+}$ we limit
ourselves to the subspace of t$_{2g}$ states,
setting aside the e$_g$ states, justified by the large octahedral crystal field
splitting (10Dq $\sim$ 1-10 eV), which for the 5d$^5$ configuration of Ir$^{4+}$
produces a single hole in the t$_{2g}$ states, and the hierarchy of
energy-scales at play, $\Delta \ll \zeta \ll 10$Dq. Iridates can thus
be thought of as one-hole t$_{2g}$ systems: dealing with one-particle systems
greatly simplifies the calculations, as particle-particle interactions are
zeroed, and expressions for one-particle ground and excited states wavefunctions
are readily derived. Resonant X-ray scattering amplitudes are then calculated
considering intra-ion transitions.
The assumption of considering the subspace spanned by the t$_{2g}$ states only 
is further justified by the observation in Ir L$_3$ edge RXMS and RIXS that the
magnetic elastic and magnetic and spin-orbit excitations resonate at $\sim10$Dq
lower energy than the main absorption line\cite{Boseggia2012,Liu2012},
indicating that they originate from initial $2p\rightarrow5d$ transitions
into the same unoccupied states within the Ir t$_{2g}$ manifold\cite{Liu2012}. 

\begin{figure}
	\centering
		\includegraphics[width=.95\columnwidth]{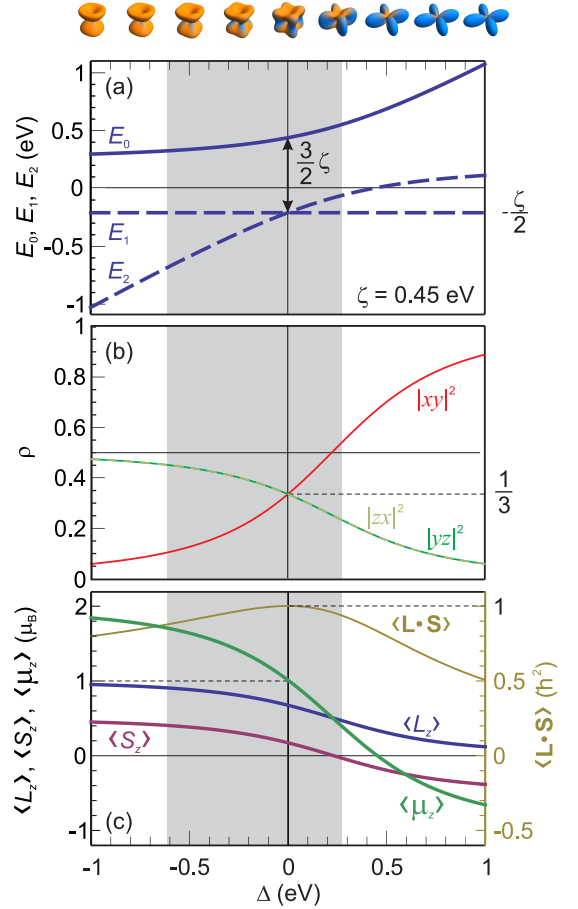}
\caption{$\Delta$ dependence of (a) eigenvalues of Eq.(\ref{hamilt}) (blue
lines), (b) the groundstate orbital occupancies of the $|xy,-\rangle$ (red),
$|yz,+\rangle$ (green) and $|zx,+\rangle$ (yellow line) states, and (c) the
expectation values of the orbital ($\langle L_z \rangle$, blue), spin ($\langle
S_z \rangle$, purple) and total ($\langle \mu_z \rangle$, green) magnetic moment
components along $z$, and expectation value of the angular part of the spin
orbit coupling ($\langle \mathbf{L} \cdot \mathbf{S} \rangle$, yellow line). The
continuous blue line in panel (a) represent the ground state energy in the
\emph{hole} representation. The corresponding wavefunction (according to
Eq.(\ref{gs_wavefunction1})) is represented at the top: blue and orange
represent the contributions of the $|xy,-\rangle$ and $\left(
|yz,+\rangle-\imath | zx,+\rangle \right)/\sqrt{2}$ states, respectively. The
shaded area in all panels represents the range of $\Delta$ values, for which
the L$_2$/L$_3$ RXMS intensity ratio is smaller than 0.1 (see
Fig.\ref{fig2}).}\label{fig1}
\end{figure}

The Hamiltonian acting on the $5d$ $t_{2g}$ states relevant to irididate
perovskites 
is written as\cite{Ament2011a,Liu2012,Ohgushi2013,Hozoi2012}
\begin{equation}\label{hamilt}
 \mathcal{H} = \zeta \mathbf{L}\cdot\mathbf{S}-\Delta \langle L_z \rangle ^2.
\end{equation}
For negligible SOC ($\zeta=0$), its eigenstates are the familiar
$|xy,\pm\rangle$, $|yz,\pm\rangle$ and $|zx,\pm\rangle$ orbitals, where
$\pm$ refers to the spin. 
In the case of iridium, however, SOC can be as large as 0.45
eV\cite{Andlauer1976}, and therefore cannot be neglected. 
For negligible tetragonal crystal-field,  i.e. for $\Delta=0$, the groundstate 
of the system is the so-called $| j_{\mathrm{eff}}=1/2\rangle$ state described
below. At intermediate couplings, the eigenstates of $\mathcal{H}$ are three 
Kramers doublets, which we write as $|0,\pm\rangle$, $|1,\pm\rangle$ and
$|2,\pm\rangle$.

An essential pre-requisite for calculating the resonant X-ray scattering
amplitudes is
to determine the eigenvalues and eigenfunctions of Eq.(\ref{hamilt}), which
for completeness we  present here. The eigenvalues (see
Supplementary Materials)
are shown in Fig.\ref{fig1}(a) for $\zeta=0.45$ eV (as extracted from
experiments\cite{Liu2012,Bogdanov2012,Gretarsson2013}) and realistic values of
$\Delta$, i.e $|\Delta|<1$ eV\cite{Liu2012,Gretarsson2013,Hozoi2012}. With five
electron filling the three doublets, one hole is left in the, say, $|0,-\rangle$
state, which is therefore the ground state of the system in the \emph{hole}
representation. The corresponding wavefunction is written as
\begin{equation}
 |0,-\rangle_{c} = \frac{C_0 |xy,-\rangle + |yz,+\rangle
- \imath |zx,+\rangle}{\sqrt{2+C_0^2}}  \label{gs_wavefunction1}
\end{equation}
for $\boldsymbol\mu\parallel(001)$, and
\begin{equation}
 |0,-\rangle_{ab} = \frac{C_0 \left( |xy,-\rangle
-\imath |xy,+\rangle\right)/\sqrt{2}+ |yz,+\rangle
+ \imath |zx,-\rangle}{\sqrt{2+C_0^2}} \label{gs_wavefunction2}
\end{equation}
for $\boldsymbol\mu\parallel(110)$, respectively, where
$2C_0=\delta-1+\sqrt{9+\delta(\delta-2)}$ and $\delta=2\Delta/\zeta$. We mostly
focus on $\boldsymbol\mu\parallel(001)$ and $\boldsymbol\mu\parallel(110)$, as
these are the cases for Sr$_3$Ir$_2$O$_7$ and A$_2$IrO$_4$ (A=Sr and Ba),
respectively. For
$\Delta=\delta=0$ ($C_0=1$), the $j_{\mathrm{eff}}=1/2$ ground state
is realized, and Eqs.(\ref{gs_wavefunction1}) and (\ref{gs_wavefunction2})
reduce to
\begin{equation}
  |j_{\mathrm{eff}}=\frac{1}{2}\rangle_{c} =  \frac{|
xy,-\rangle +|yz,+\rangle-\imath | zx,+\rangle}{\sqrt{3}} \label{gs001}
\end{equation}
and
\begin{equation}
  |j_{\mathrm{eff}}=\frac{1}{2}\rangle_{ab} = 
\frac{\left(|xy,-\rangle-\imath|xy,+\rangle\right)/\sqrt{2} +|yz,+\rangle+\imath
|zx,-\rangle}{\sqrt{3}} \label{gs110}
\end{equation}
respectively. It has to be stressed here that the expression of
$|j_{\mathrm{eff}}=1/2\rangle$ is different in the
two cases $\boldsymbol\mu\parallel(001)$ and $\boldsymbol\mu\parallel(110)$.

At the top of Fig.\ref{fig1}, a real-space representation is given of
$|0,-\rangle$
as a function of $\Delta$: the well-known ``cubic'' shape
of the $|j_{\mathrm{eff}}=1/2\rangle$ wavefunction is evident for $\Delta=0$. At
finite values of $\Delta$,  the admixture of orbital contributions
changes: in particular, in the limit for $\Delta\gg\zeta$, the ground state
reduces to the $|xy,-\rangle$, while it reads $\left(|yz,+\rangle-\imath |
zx,+\rangle \right)/\sqrt{2}$ for $\Delta\ll -\zeta$. This is also seen in
Fig.\ref{fig1}(b), where the relative orbital occupancy is shown: this is the
same for the three orbitals, i.e. $\left(1/\sqrt{3}\right)^2=1/3$, at
$\Delta=0$. Fig.\ref{fig1}(c), finally, shows the $\Delta$ dependence of the
expectation values of the orbital ($\langle L_z \rangle$, blue), spin ($\langle
S_z \rangle$, purple) and total ($\langle \mu_z \rangle$, green) magnetic moment
components along $z$, for $\boldsymbol\mu\parallel(001)$. Note that
$\langle\mu_z\rangle=1$ for $\Delta = 0$ and $\langle\mu_z\rangle=0$ for $\Delta
= \zeta$. The $\Delta$ dependence of the expectation value of the spin orbit
coupling operator ($\langle \mathbf{L} \cdot \mathbf{S} \rangle$, yellow line),
independent of the magnetic moment orientation, is also shown to reach a
maximum of 1 at $\Delta = 0$, as expected.  (See Supplementary Materials for the
expression of the expectation values of the momentum and spin-orbit operators
for both $\boldsymbol\mu\parallel(001)$ and $(110)$).

Having obtained the eigenvalues and eigenfunctions of Eq.(\ref{hamilt}) 
we now proceed to the main task of calculating the required resonant X-ray
scattering amplitudes. RIXS is a second-order process described by the
Kramers-Heisenberg (KH) formula: 
\begin{equation}\label{FF}
    \mathcal{A}_{|f,\pm\rangle}^{\boldsymbol\epsilon \boldsymbol\epsilon^\prime}
= \sum_{n}
\frac{\langle f,\pm|\mathcal{D}_{\boldsymbol\epsilon^\prime}^{\dagger}|n\rangle
    \langle n|\mathcal{D}_{\boldsymbol\epsilon}|0,-\rangle
    }{E_0-E_n+\hbar\omega+\imath\Gamma_n}
\end{equation}
is the scattering amplitude from the ground state, $|0,-\rangle$ (of energy
$E_0$) to the final states $|f,\pm\rangle$ ($f=0,1,2$, energy $E_f$). $n$
runs over all the intermediate states of energy $E_n$ and intrinsic linewidth
$\Gamma_n$. $\mathcal{D}_{\boldsymbol\epsilon}$
($\mathcal{D}_{\boldsymbol\epsilon^\prime}^\dagger$) is the absorption
(emission) transition operator, where 
$\boldsymbol\epsilon$ ($\boldsymbol\epsilon^\prime$) defines the polarisation of
the incoming (outgoing) photons. At resonance ($\hbar\omega\approx E_0-E_n$),
this is the leading term in the RIXS cross-section and the only one considered
here. Assuming that, at a given edge ($E_n=E$), the intermediate states have all
the same intrinsic linewidth $\Gamma_n=\Gamma$, the denominator of Eq.(\ref{FF})
can be discarded, and the expression of the atomic form factor simplifies to
$\mathcal{A}_{|f,\pm\rangle}^{\boldsymbol\epsilon
\boldsymbol\epsilon^\prime}\propto\sum_{n}\langle f,\pm|\mathcal{D}_{
\boldsymbol\epsilon^\prime}^{\dagger}|n\rangle
\langle n|\mathcal{D}_{\boldsymbol\epsilon}|0,-\rangle$. 

To calculate the matrix elements of the resonant scattering amplitudes, we use
the atomic wavefunctions derived within the single ion model, and restrict
ourselves to the case of dipolar transitions. The scattering geometry (sketched
in the Supplementary Materials) is defined through the azimuthal, $\theta$
($\theta^{\prime}$), and polar, $\phi$ ($\phi^{\prime}$), angles of the incident
(scattered) photon wavevector, $\mathbf{k}$ ($\mathbf{k}^\prime$), in the sample
reference system. The polarisation $\boldsymbol\epsilon$
($\boldsymbol\epsilon^\prime$) of the incident (scattered) photon is projected
on a two-vector basis, perpendicular ($\sigma$) and parallel ($\pi$) to the
scattering plane.

\begin{figure}
	\centering
		\includegraphics[width=\columnwidth]{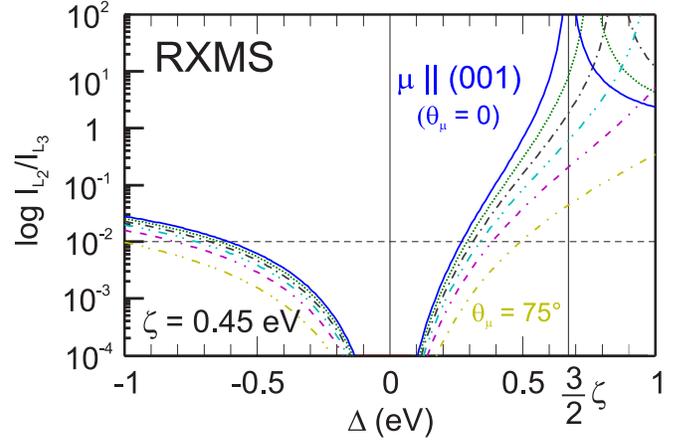}
\caption{L$_2$/L$_3$ RXMS intensity ratio (logarithmic scale) as a
function of the tetragonal crystal field splitting $\Delta$ ranging from -1 to 1
eV, for a given value of the SOC constant ($\zeta=0.45$ eV). Different line
styles correspond to values of $\theta_\mu$ from 0 to 90$^\circ$ in steps of
15$^\circ$. }\label{fig2}
\end{figure}

The resonant elastic X-ray scattering amplitude (REXS)  is 
obtained in the special case that $|f,\pm\rangle\equiv|0,-\rangle$. 
For a crystal, the REXS cross-section in general is proportional to
$|\mathcal{F}^{\boldsymbol\epsilon \boldsymbol\epsilon^\prime}
\left(\mathbf{Q}\right)|^2$, where $\mathcal{F}^{\boldsymbol\epsilon
\boldsymbol\epsilon^\prime} \left(\mathbf{Q}\right)$ is the unit cell structure
factor, and $\mathbf{Q}=\mathbf{k}^\prime-\mathbf{k}$. For the specific case of
antiferrromagnetic order considered here, 
the RXMS structure factor is derived as a sum over two 
sub lattices ($A$ and $B$, say), so that
\begin{equation} \label{REXS}
 \mathcal{F}^{\boldsymbol\epsilon \boldsymbol\epsilon^\prime}
\left(\mathbf{Q}\right) = 
 f_A^{\boldsymbol\epsilon \boldsymbol\epsilon^\prime} \sum_A e^{\imath
\mathbf{Q}\cdot\mathbf{r}_A} +
f_B^{\boldsymbol\epsilon \boldsymbol\epsilon^\prime} \sum_B e^{\imath
\mathbf{Q}\cdot\mathbf{r}_B},
\end{equation}
with $f_A^{\boldsymbol\epsilon
\boldsymbol\epsilon^\prime}=\mathcal{A}_{|0,-\rangle}^{ \boldsymbol\epsilon
\boldsymbol\epsilon^\prime}=-f_B^{\boldsymbol\epsilon
\boldsymbol\epsilon^\prime}$, $\mathbf{r}_A$ ($\mathbf{r}_B$) the
position of the $A$ ($B$) atom within the magnetic unit cell, and
$\mathbf{Q}=\mathbf{Q}_{AF}$, the
antiferromagnetic propagation wavevector. 

We now consider the RXMS intensity branching ratio, 
$\left|\mathcal{F}^{\sigma\pi}_{L_2}\right|^2/\left|\mathcal{F}^{\sigma\pi
}_{L_3}\right|^2$, as this is the quantity, readily measured in experiments,
which has been used to
infer the existence of a $j_{\mathrm{eff}}=1/2$ ground state in various iridate
perovskites.
With $\boldsymbol\mu\parallel(001)$ the REXS scattering amplitudes at the L$_2$
edge are
given by $\mathcal{A}_{|0,-\rangle}^{\sigma\pi}=\imath \left(C_0-1
\right)^2 \cos \theta^{\prime}/\left(2+C_0^2\right)$
and $\mathcal{A}_{|0,-\rangle}^{\pi\sigma}=-\imath \left(C_0-1
\right)^2 \cos \theta/\left(2+C_0^2\right)$, while at the L$_3$ edge these
read $\mathcal{A}_{|0,-\rangle}^{\sigma\pi}=- \imath  \left[C_0(C_0-2)-2
\right]\cos \theta^{\prime}/\left(2+C_0^2\right)$ and $
\mathcal{A}_{|0,-\rangle}^{\pi\sigma}=\imath \left[C_0(C_0-2)-2
\right] \cos \theta/\left(2+C_0^2\right)$. Given the scattering amplitudes and
the atomic positions within the unit cell, the RXMS intensity branching ratio
is given by 
\begin{equation}\label{L23ratio}
\frac{\left|\mathcal{F}^{\sigma\pi}_{L_2}\right|^2}{\left|\mathcal{F}^{\sigma\pi
}_{L_3}\right|^2}=\frac{\left|\mathcal{F}^{\pi\sigma}_{L_2}\right|^2}{
\left|\mathcal{F}^{\pi\sigma }_{L_3}\right|^2} =\frac { (C_0-1)^4}{\left
[C_0(C_0-1)-2\right]^2}
\end{equation}
Its dependence on the tetragonal distortion is shown in  Fig.\ref{fig2} (blue
curve) for
$\zeta=0.45$ eV, and is consistent with previous calculations with 
$\boldsymbol\mu\parallel(001)$ \cite{Kim2012}, relevant to the case of
Sr$_3$Ir$_2$O$_7$. 
The calculated branching ratio drops to zero for $\Delta=0$, while it diverges
for $\Delta=3\zeta/2$. In the limit for $\Delta\gg\zeta$, the ratio tends to
unity, and to 1/4 for $\Delta\ll -\zeta$. It was claimed that the experimental
ratio of at most 1\% provides the lower and upper bounds for nearly pure
$j_{\mathrm{eff}}=1/2$ ground state. We note, however, that these bounds
correspond to a relatively large energy window in $\Delta$ (-0.61 eV $<\Delta<$
0.27 eV), for which the ground state may deviate considerably from the pure
$j_{\mathrm{eff}}=1/2$ state, as seen in the substantial change of the shape of
the ground state wavefunction, of the orbital occupancy ($0.1<|xy|^2<0.54$), and
of the expectation values of $\langle L_z \rangle$, $\langle S_z \rangle$ and
$\langle \mathbf{L} \cdot \mathbf{S} \rangle$ (Fig.\ref{fig1}).
Nonetheless, the experimentally observed branching ratio in Sr$_3$Ir$_2$O$_7$ is
less than 1\%,
and it is probably reasonable to conclude that in this case the RXMS
experiments 
provide evidence of a $j_{\mathrm{eff}}=1/2$ ground state.

Figure \ref{fig2} also shows the dependence of the RXMS branching ratio on the
direction of $\boldsymbol\mu$, defined through the $\theta_\mu$
angle ($\boldsymbol\mu\parallel (001)$ for $\theta_\mu=0$, while
$\boldsymbol\mu\parallel (110)$ for $\theta_\mu=90^\circ$).
When
$\boldsymbol\mu$ is progressively lowered into the basal plane, the divergence
in the ratio moves towards higher values of $\Delta$, and
eventually disappears for $\theta_\mu=90^\circ$. Consequently, for magnetic
moments lying in the \emph{ab}-plane, the L$_2$ edge RXMS intensity is
identically zero, \emph{irrespective} of the tetragonal crystal field splitting
of the t$_{2g}$ states. (This can be traced to the fact that for
$\boldsymbol\mu\parallel (110)$,
$\mathcal{A}_{|0,-\rangle}^{\sigma\pi}=\mathcal{A}_{|0,-\rangle}^{ \pi\sigma}
\equiv 0$, a consequence of the particular coherent superposition of  states in 
Eq.(\ref{gs110})). This is an important result as it implies that the
$j_{\mathrm{eff}}=1/2$ ground state cannot be inferred
by the L$_2$/L$_3$ RXMS intensity ratio when
$\boldsymbol\mu\bot (110)$, as has been claimed in the case of A$_2$IrO$_4$
(A=Sr,
Ba) compounds\cite{Kim2009,Boseggia2013}. 

\begin{figure}
	\centering
		\includegraphics[width=\columnwidth]{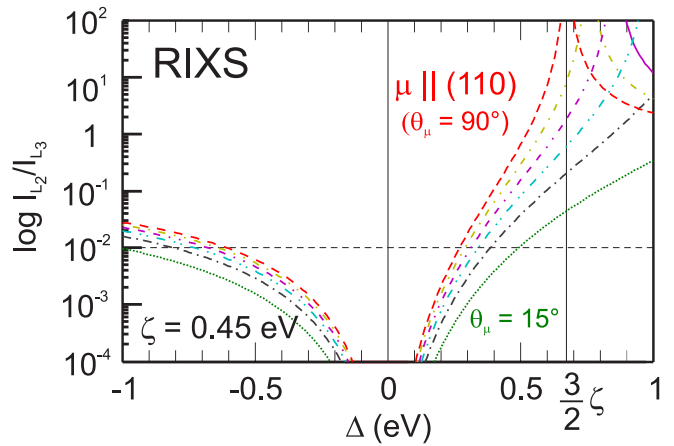}
\caption{Ratio of the L$_2$/L$_3$ ``spin-flip'' intensity ratio (logarithmic
scale) as a function of the tetragonal crystal field splitting $\Delta$ ranging
from -1 to 1 eV, for a given value of the SOC constant
($\zeta=0.45$ eV). Different line
styles correspond to values of $\theta_\mu$ from 90 to 0$^\circ$ in steps of
15$^\circ$. }\label{fig3}
\end{figure}

We have also calculated the RIXS amplitudes associated with transitions to
excited
states within the 5d t$_{2g}$ manifold. Here we focus on the ``spin-flip''
channel from
the $|0,-\rangle$ ground state to the $|0,+\rangle$ final state, pertinent 
to the interpretation of experiments that have successfully observed 
magnons. In Fig.\ref{fig3} we report the L$_2$/L$_3$ ``spin-flip'' intensity
ratio as a
function of $\Delta$ (for $\zeta=0.45 $ eV). Remarkably, it is seen that the
dependence on
the direction of $\boldsymbol\mu$ is opposite to that of RXMS: the ratio is
identically zero for magnetic moments along the $c$-axis, irrespective of the
tetragonal crystal field splitting, while for all other $\boldsymbol\mu$
orientations it drops to zero only for $\Delta=0$, i.e. when the
$j_{\mathrm{eff}}=1/2$ ground state is realized. 

It remains to consider the extent to which the salient results of our
calculations may be 
altered by the inclusion of additional effects, such as the e$_g$ states,
electronic band formation, many-body interactions, etc. Although these will all
no doubt 
affect the quantitative dependences on $\theta_\mu$ and $\Delta$ shown in Figs.\
\ref{fig2}
and \ref{fig3}, we nevertheless expect the qualitative features of our
results to remain unchanged. The reason is that ultimately, effects such as the
extinguishing of the RXMS L$_2$ intensity for $\theta_\mu=90^\circ$, depends on
the symmetry of the 5d wave function which partially persists into the solid.
When comparing with the results of RXMS experiments, it should be appreciated
that 
with the limited energy resolution usually employed ($\sim$1eV) what is actually measured 
is the sum of elastic plus partially integrated inelastic responses. Thus the differences 
exhibited by the Sr and Ba ``214'' compounds $-$ the L$_2$ intensity being small and finite in the
former and zero in the latter $-$ could be related to the detailed differences of the excitation spectra
for the two systems.

In conclusion, we have developed a single-ion model relevant to the
iridate perovskites by which we are able to understand how the results of
resonant X-ray elastic and inelastic scattering experiments relate to
their underlying electronic structure. Our model treats the SOC, $\zeta$,
and tetragonal crystal field, $\Delta$,  on an equal footing and,
most importantly, explicitly takes into account the direction $\theta_\mu$ of
the Ir$^{4+}$ magnetic moments, $\mu$.  The results of our calculations reveal
the full complexity  of the relationship between $\zeta$, $\Delta$ and
$\theta_\mu$ in determining the RXMS and RIXS cross-sections at the L$_2$ and
L$_3$ edges. In terms of existing experimental
data\cite{Kim2009,Boseggia2013,Ohgushi2013}, our results clearly call for a
reinterpretation as the L$_2$  RXMS cross-section is zero for
$\theta_\mu=90^\circ$ irrespective of the value of $\Delta$. Our calculations
will also be useful in guiding future experiments, including, for example, the
properties of thin films of iridates have started to be
explored\cite{Serrao2013,Nichols2013}. Here the strong magneto-elastic coupling
inherent to the iridates might allow epitaxial strain effects to be used to
control  magnetic structure,  including moment reorientation transitions, etc.
According  to our calculations, any such transitions will have a  clear
signature in the RXMS and RIXS intensity ratios. Finally, although the focus
here has been on iridate perovskites, the general approach we have developed can
be readily extended to other systems of interest. 

\emph{Acknowledgments} - The authors would like to acknowledge F. de
Bergevin, L. Braicovich, G. Ghiringhelli and C. Mazzoli for stimulating and
elucidating discussions. The work in the UK was supported through a grant from
the EPSRC, and in Denmark by the Nordea Fonden and the Otto M\o nsteds Fond.  

\bibliography{biblio}

\end{document}